# DNA-MATRIX: a tool for DNA motif discovery and weight matrix construction

**Chandra Prakash Singh [1]**
Department of Computer Sciences,
R.S.M.T., U.P. College,
Varanasi (India)

**Sanjay Kumar Singh[2]**
Department of Computer Sciences,
R.S.M.T., U.P. College,
Varanasi (India)

**Dr. Feroz Khan[3]***
Metabolic & Structural Biology Division,
Central Institute of Medicinal & Aromatic Plants (CSIR),
Lucknow (India)

**Prof. Durg Singh Chauhan[4]**
Uttrakhand Technical University
Dehradun (India)

**Abstract—** *In computational molecular biology, gene regulatory binding sites prediction in whole genome remains a challenge for the researchers. Now a days, the genome wide regulatory binding site prediction tools required either direct pattern sequence or weight matrix. Although there are known transcription factor binding sites databases available for genome wide prediction but no tool is available which can construct different weight matrices as per need of user or tools available for large data set scanning by first aligning the input upstream or promoter sequences and than construct the matrices in different level and file format. Considering this, we developed a DNA-MATRIX tool for searching putative regulatory binding sites in gene upstream sequences. This tool uses the simple biological rule based heuristic algorithm for weight matrix construction, which can be transformed into different formats after motif alignment and therefore provides the possibility to identify the most potential conserved binding sites in the regulated genes. The user may construct and save specific weight or frequency matrices in different form and file formats based on user based selection of conserved aligned block of short sequences ranges from 6 to 20 base pairs and prior nucleotide frequency before weight scoring.*

**\* Corresponding author Email: f.khan@cimap.res.in**

## INTRODUCTION

The most important functional element in any genome is the transcription factor (TF) and the sites within the DNA to which they bind called as transcription factor binding sites (TFBS). Despite considerable efforts to date, these sites identification remains a challenge in computational molecular biology. Researchers have taken different approaches in developing motif discovery tools and the progress made in this area of research is very encouraging. TFBS are usually short (~ 5-15 base pair [bp]) and they are frequent in degenerate sequence motifs. Although degenerate consensus sequences are frequently used to depict the binding specificities of TFs, they do not contain precise information about the relative likelihood of observing the alternate nucleotides at the various positions of a TFBS. A common way of representing the degenerate sequence preferences of a DNA-binding protein is by a position weight matrix, also known as a position specific scoring matrix (PSSM). The elements of PSSM correspond to scores reflecting the likelihood of a particular nucleotide at a particular position. Recent advances in genome sequence availability and in high–throughput gene expression analysis methods have allowed for the development of computational methods for regulatory motif finding. As a result, a large number of motif finding algorithms have been implemented. So far no such tool is available to construct the user based weight matrices at one place through non-aligned input non-coding sequences. Earlier algorithms [9,11,20] use promoter sequences of co regulated genes from single genome and search for statistically overrepresented motifs. Recent algorithms are designed to use phylogenetic footprints or orthologous sequence set. All the tools made somehow predict the 50 to 80% correct binding sites but at the same time they predict the false positives or in other words we can say hypothetical novel binding sites, which may or may not be a real binding site. Since we mostly compare the predictions based on available database size of transcription factor binding sites, but at the same time, if we think biologically, all the binding sites are not reported or discovered so far. Thus, it might be possible to identify or discover in future after completion of sequencing projects of most of the diverse family organisms. It takes some time but predictions are going on, therefore we cannot ignore all these predictions or motif discoveries at present, which might be true in future. Based on it we developed a tool for motif discovery from non-aligned sequences, but it should be coregulated or coexpressed or orthologous. This is because all the evolutionary functional sites remain unchanged during the evolution. Thus our tool predict the conserved blocks based on multiple sequence alignment and show the local regions which are similar in nature and accordingly score the block based on multiple alignment scoring scheme. Now users can chose the functionally conserved blocks and predict the alignment or weight matrix in different formats and also get the logo representation of conserved motifs. Even user can predict the exact binding site width or length of nucleotides which are necessary for regulatory protein or transcription factor binding sites. These features are not so far covered in any available web tools and therefore DNA-MATRIX tool revealed the biological highlight of unexplored molecular phenomenon as compared to other tools which are mostly some bound with algorithm or machine learning methods. These approaches in biological system are not upto the mark and therefore

.

integrated approaches are used to understand the complex biological processes [1,4,5]. It is now evident that most of the motif finding or discovery tools have been shown to work successfully in lower organisms, but perform significantly worse in higher organisms [2,6,7,10]. Over the past few years, numerous tools have become available for the prediction of gene regulatory binding sites as in [18, 23]. Especially popular are those webtools which uses known binding sites information collected in databases such as TRANSFAC [13,], ooTFD [25], EpoDB [17], TRANSCompel [10]. By using only limited data for motif prediction, it seems biased approach e.g, D-Matrix [22], TFsitescan [25], MATCH$^{TM}$, TESS [17], AliBaba2 [8], SIGNAL SCAN [15], MATRIX SEARCH [3], MatInspector [16], Fuzzy clustering tool [14], FUNSITE [10], Gibbs Sampling [12] and other tools available on the web. Therefore to overcome this known binding sites biasness, we have developed a tool called DNA-MATRIX which is a motif discovery as well as weight matrix construction tool based on non-aligned sequences and user friendly. One can chose any conserved block in the aligned input sequences for the development of weight matrix to search genome wide level similar sites or motifs through available tools such as PoSSuMsearch and/or RSAT-Patser.

## ALGORITHM

DNA-MATRIX takes non-aligned DNA sequences as input and predicts the conserved blocks of different lengths, depending upon user expertise, one can select the appropriate block and then predict the frequency of each nucleotides at given position. This finally forms the matrix (n x m), where 'n' refers four types of nucleotides i.e., ATGC and 'm' refers length of the block or putative binding site (motif). Based on nucleotide repetition, DNA-MATRIX can predict the frequency matrix, alignment matrix and weight matrix along with motif signature and logo representation. Also give the degenerate consensus sequence according to IUPAC/IUB convention as output results.

Scoring of the weight matrix was done through following equation:

$$Weight_{i,j} = \ln (n_{i,j}+p_i)/(N+1) \div p_i$$

$$Weight_{i,j} = \sim \ln f_{i,j} / p_i$$

N - Total number of sequences
$n_{i,j}$ - number of times nucleotide $i$ was observed in position $j$ of the alignment
$f_{i,j} = n_{i,j}/N$ = frequency of letter $i$ at position $j$
$p_i$ - a priori probability of letter $i$

Positive weight $i,j$ means that frequency letter $i$ at position $j$ of the alignment is higher than a priori probability of this letter. Weight can be interpreted as an estimate of the free energy of the protein binding to this site. Informational content is higher for the alignment whose letter frequencies most differ from the a priori probabilities. Informational content is a measure of discrimination between the binding of a functional DNA sequence and an arbitrary DNA sequences.

The multiple sequence alignment method was implemented in the tool, which uses progressive dynamic programming algorithm for local alignments in related sequences in order to detect short conserved regions of motifs that may not be in the same positions. The acceptable matrix or block length is 6 to 25 bp based on reported known TF binding sites [25], which allowed the detection of both short and longer conserved motifs within the input sequences. Developed matrix represent the nucleotide conservation in each position of the motif and can be used to search for genome wide prediction of unexplored functional sites in non-coding genomic sequences by using RSAT-Patser [19, 24] and PoSSuMsearch [22] tools. Besides, user can transformed the matrices in to different file formats, depending upon the other software requirement or user need. Moreover, user can also change the matrix or conserved motif sequences in to raw input file format used for logo representation of nucleotides in each position through WebLogo program [21].

## IMPLIMENTATION

The software is developed in C and the program is wrapped by a Perl script. The top panel is used to paste the input sequences (or non-aligned) and to specify the name and width of motif (optional) to be search. The results panel contains few major sections: alignment blocks, consensus pattern/motif, frequency matrix, alignment matrix, prior frequency selection, weight matrix and signature sequence as per IUPAC code and logo representation. DNA-MATRIX can use both orthologous/co-regulated genes upstream sequences as well as unknown multiple genomic sequences (minimum 6 sequence and maximum no limit, but length should not exceed more than 400 bp in length for each input sequence). For testing, we used the unknown or random sequences as negative data set and also used known PurR transcription factor binding sites as positive data set for evaluation of tool prediction performance. Results showed that DNA-MATRIX successfully predict the true positive hits of *Escherichia coli* PurR transcription factor binding sites. Predictions performance showed 90% of accuracy.

### Merits of DNA-MATRIX tool

DNA-MATRIX tool is comparable and much better than existing D-Matrix tool available at CIMAP website network domain (www.cimap.res.in) [22], because limitation of D-Matrix tool is not to align the input short sequences and only consider known motifs for weight matrix construction. It also not included all the prior nucleotide frequencies of A/T:G/C ratios such as, 0.3:0.2, 0.2:0.3 and 0.4:0.1. It only covers 0.3:0.2 prior nucleotide frequencies on the basis of available bacterial genomes, which showed this type of frequency and so exclude others. Following so, these limitations were overcome in the DNA-MATRIX tool, which first align the input upstream sequences (unknown) by using multiple sequence alignment algorithm such as progressive dynamic programming used in ClustalW program (www.ebi.ac.uk/clustalw/) and other motif discovery tools and then on the basis of nucleotide conservation blocks, the rule based heuristic algorithm scored it and select the maximum score block of sequences. Later these highly conserved block (as selected by user) used for frequency matrix development and finally converted into weight matrix through standard scoring system as mentioned in the methodology section. Besides, we have also given option to user to select the appropriate prior nucleotide frequency before weight matrix generation in different formats.

## CONCLUSION

The DNA-MATRIX tool predict the unexplored novel binding sites or motifs based on biological principle of evolutionary conservation in functional sites, which support the appropriate theoretical concept for tool development. Motif prediction in whole genome or large data set remains a challenge for the researchers because of the complex process of regulatory genomics, since it is still poorly understood. DNA-MATRIX uses the simple biological rule based algorithm or heuristic approach used in multiple sequences alignment algorithms, to search the conserved sites and later predict the weight matrix. These can be changed in to different file format as per need of user expertise and problem taken under study. This tool can be used for any DNA genomic sequences for motif discovery in short sequences and later applied to predict in large dataset such as genome level. We have covered all three types of nucleotide prior frequencies i.e., A/T:G/C ratio in the range of 0.3:0.2, 0.2:0.3 and 0.4:0.1 in the tool, while deriving weight matrix, keeping in mind all sorts of bacterial genomic nucleotide frequencies reported so far. This will result in identification of great variety of different conserved motifs which might be regulatory in nature but reported yet. The user may construct and save specific weight or frequency matrices in different formats derived through user selected set of block or sequences.

## ACKNOWLEDGEMENT

Author is grateful to the technical reviewers for the comments, which improved the clarity and presentation of the paper. Author wishes to thank Mr. Naresh Sen, Project Assistant, Bioinformatics Centre, Biotechnology Division, CIMAP, Lucknow, for all the discussions and contributions during the Software development. We acknowledge the Central Institute of Medicinal & Aromatic Plants (Council of Scientific & Industrial Research), Lucknow for providing me facility and all support for completion of my Ph.D research work.

## REFERENCES

[1] Bhcher P. (1999). Regulatory elements and expression profiles. Curr.Opin Struct. Biol., 9:4000-407.

[2] Chandra Prakash Singh, Feroz Khan, Bhartendu Nath Mishra, Durg Singh Chauhan (2008). Performance evaluation of DNA motif discovery programs. Bioinformation 3(5): 205-212.

[3] Chen Q.k., Hertz G.Z. and Stormo G.D. (1995). MATRIXSEARCH 1.0: a computer program that scans DNA sequences for transcriptional elements using a database of weight matrices. Comput. Appl. Biosci., 11:563-566.

[4] Conkright M.D.Guzman E. Flechner L.Su. A.I. Hogenesch J.B. and Montminy M. (2003). Genome-wide analysis of CREB target genes revealsa core promoter requirement for cAMP responsiveness. Mol. Cell, 1101 –1108.

[5] Fessele S., Maier H., Zischek C., Nelson P.J. and Wernere T. (2002). Regulatory context is a crucial part of gene function Trends Genet., 18:60-63.

[6] Fickett J.W.and Wasserman W.W. (2000). Discovery and modeling of transcriptional regulatory regions .Curr. Opin. Biotechnol., 11:19-24.

[7] Gobling E. Kel_Margulis O.V., Kel A.E. and Wingender E. (2001). MATCH-TM A tool for searching transcription factor binding sites in DNA sequences. Application for the analysis of human chromosomes. proceedings of the German Conference on Bioinformatics GCB'01 Braunschweig, Germany , October 7-10, pp. 158-161.

[8] Grabe N. (2000). AlibBaba2: context specific identification of transcription factor binding sites. In Silico Biol., 1:0019.

[9] Kel A. Kel –Margoulis O. Borlack J., Tchekmenev D.And Wingender E. (2005). Databases and tools for in silico analysis of regulation of gene expression.In Borlak J. (ed.) ,Handrakhin of Toxicogenomics VCH Weinheim, pp. 253-290.

[10] Kel A.E. kondrakhin Y.V. ,Kolpakov Ph.A. Kel O.V. Romashenko A.G.,Wingender E., Milanesi L. And Kolchanov N.A. (1995). Compuater tool FUNSITE for analysis of eukaryotic regulatry genomic sequences. proc. Int. Conf. Intell. Syst. Mol. Bip., 3:197-205.

[11] Kel –Margoulis O.V., Kel A.E. Reuter I., Deineko I.V. and Wingender E. (2002). TRANS Compel: a database on comoposite regulatory elements in eukaryotic genes. Nucleic Acids Res., 30:332-334.

[12] Lawrence CE, Altschul SF, Boguski MS, Liu JS, Neuwald AF, Wootton JC (1993). Detecting subtle sequence signals: a Gibbs sampling strategy for multiple alignment. Science, 262 (5131):208-214.

[13] Matys V. Fricke E. Geffers R. Gobling E., Haubrock M., Hehl R Hornischer K., Karas D., Kel A.E. Kel Margoulis O.V.et al . (2003). TRANSFAC: transcriptional regulation, from patterns to profiles .Nucleic Acids Res., 31,374-378.

[14] Pickert L., Reuter I., Klawonn F. and Wingender E.(1998). Transcription regulatory region analysis using signal detection and fuzzy clustering. Bioinformatics, 14:244-251.

[15] Prestridge D.S (1996). SIGNAL SCAN 4.0: additional databases and sequence formats. Comput. Appl. Biosci., 12:157-160.

[16] Quandt K., Frech K., Karas H., Wingender E. and Werner T. (1995). MatInd and MatInstpector: new fast and versatile tools for detection fo consensus matches in nucleoitide sequence data. Nucleic Acids Res., 23:4878-4884.

[17] Stoeckert C.J. Jr Salas F., Brunk B. and Overton G.C. (1999). EpoDB: a prototype database for the analysis of genes expressed during vurtebrate erythropoiesis. Nucleic Acids Res., 27:200-203.

[18] Stormo G.D. (2000). DNA Binding sites: representation and discovery. Bioinformatics, 16:16-23.

[19] RSAT (Patser/Consensus): Regulatory Sequence Analysis Tool (http://embnet.cifn.unam.mx/rsa-tools).

[20] Hertz G.Z., Stormok G.D. ( 1999). Identifying DNA and protein patterns with statistically signficant aligaments of multiple sequences. Bioinformatics, 15:563-577.

[21] Web Logo (htt:/weblogo. berle;eu. edu/logo.cgis)

[22] Naresh Sen, Manoj Mishra, Abha Meena, Feroz Khana, Ashok Sharma (2009). D-MATRIX: A web tool for constructing weight matrix of conserved DNA motifs. Bioinformation, 3(10):415-418

[23] Khan F., Agarwal S., Mishra B.N. (2005). Genome wide identification of DNA binding motifs of NodD-factor in Sinorhizobium meliloti and Mesorhizobium loti. Journal of Bioinformatics and Computational Biology, 3(4):1-30.

[24] Van Helden J., Andre B., Collado-Vides J. (2000). A web site for the computational analysis of yeast regulatory sequences. Yeast, 16(2):177-187.

[25] Ghose D. (2000). Object-orinnted Transcription factor binding database (ooTFD). Nucleic Acid Res., 28(1):308-310.

## AUTHORS PROFILE

**1. Chandra Prakash Singh** is currently a research scholar at Uttarakhand Technical University, Dehradun, India, with the Subject of Computer Science. He completed his master degree in computer application (M.C.A.) in 2001 from the M.G.K.V. University; he has published 2 research papers in international journals and 4 research paper in national journal. He is working as a Lecturer in Deptt.of Computer Application, R.S.M.T, Varanasi. He has 9 Years of the experience in the field of research/academics.

**2. Dr. Feroz Khan** has received his M.Sc. degree in Botany from the C.S.J.M. Univ., Kanpur (2000), M.Tech. degree in Biotechnology (2003) and Ph.D. in Bioinformatics (2009) from the Uttar Pradesh Technical Univ., Lucknow. He is currently working as a Scientist-C in Metabolic and Structural Biology Division, Central Institute of Medicinal & Aromatic Plants (CSIR), Lucknow (India). His expertise in the development of in silico models to identify gene regulatory network, QSAR, ADMET analysis, molecular interaction study between ligand and target protein through docking and virtual screening of lead phytomolecules/derivatives. He has 18 international research papers in his credit published in peer reviewed and indexed journals. He is recipient of CSIR-SRF fellowship (2004-2007). He is a member of national societies, academic institute and research organizations. He is also serving as Bioinformatics Reviewer in few international journals e.g. African Journal of Biotechnology, Journal of Medicinal and Aromatic Plant Sciences, Drug Design and Discovery etc.

**3. Sanjay Kumar Singh** has completed his M. Tech (C.S.) from Uttar Pradesh Technical University, Lucknow. in 2006. He has done B.E. (C.S.) from NMU, Jalgaon, in 1998. He did PGDM with dual specialization in Marketing and IT from ITS, Ghaziabad in 2000. He received the outstanding staff award in IT from U.P.T. Univ. He is also working as a Reader in Deptt.of Computer Application, R.S.M.T, Varanasi. He has 11 Years of the experience in the field of academics/research.

**4. Prof. D.S. Chauhan** completed his B.Sc Engg.(1972) in electrical engineering at I.T. B.H.U., M.E. (1978) at R.E.C. Tiruchirapalli (Madras University) and PH.D. (1986) at IIT/Delhi. His brilliant career brought him to teaching profession at Banaras Hindu University where he was Lecturer, Reader and then has been Professor till today. He has been director KNIT Sultanpur in 1999-2000 and founder vice Chancellor of U.P.Tech. University (2000-2003-2006). Later on, he has served as Vice-Chancellor of Lovely Profession University (2006-07) and Jaypee University of Information Technology (2007-2009) Currently he has been serving as Vice-Chancellor of Uttarakhand Technical University for (2009-12) Tenure.

He has supervised 12 Ph.D., one D.Sc and currently guiding half dozen research scholars. He has authored two books and published and presented 95 research papers in international journals and international conferences and wrote more than 20 articles on various topics in national magazines. He is Fellow of institution of Engineers and Member IEEE.